The Electronic and Magnetic Properties of Magnetoresistant $Nd_{1-x}Sr_xMnAsO$ Oxyarsenides

E. J. Wildman [1], N. Emery [2] and A. C. Mclaughlin* [1]

[1] The Chemistry Department, University of Aberdeen, Meston Walk, Aberdeen, AB24 3UE, Scotland.

[2] Institut de Chimie et des Materiaux Paris Est, ICMPE/GESMAT, UMR 7182 CNRS-Universite Paris Est Creteil, CNRS 2 rue Henri Dunant, 94320 Thiais, France

* a.c.mclaughlin@abdn.ac.uk

**Abstract**

The oxypnictides $Nd_{1-x}Sr_xMnAsO$ have been successfully synthesised with *x* up to 0.1. A synchrotron X-ray diffraction study demonstrates that there is no change in crystal symmetry upon doping with Sr. An expansion of the inter-layer distance between Nd-O-Nd and As-Mn-As blocks is observed with increasing *x*. Results from variable temperature neutron diffraction and resistivity measurements show that the local moment antiferromagnetic order of the Mn spins is preserved as the $[MnAs]^-$ layers are hole doped and the materials are driven metallic for $x \geq 0.05$. A sizeable positive magnetoresistance is observed at low temperature which demonstrates that multiple MR mechanisms are possible in *Ln*MnAsO oxypnictides.

## I. INTRODUCTION

Iron oxyarsenides, LnFeAsO, with the tetragonal ZrCuSiAs structure, have recently received much attention due to the observation of high temperature superconductivity. A maximum $T_c$ of 55 K has been achieved via substitution of oxygen with fluorine [1, 2, 3] or by creating oxygen vacancies [4]. We have recently investigated the Mn analogue LnMnAsO (Ln = La, Nd) [5, 6]. Both materials are antiferromagnetic with $T_{Mn}$ ~ 360 K. Below $T_{Mn}$ the $Mn^{2+}$ moments are aligned antiferromagnetically in the ab plane, but ferromagnetically along c and the moment is ordered parallel to the c axis. At 23 K antiferromagnetic order of the Nd spins is observed with moments aligned parallel to a, which results in a spin reorientation of the Mn moments into the basal plane [6, 7]. The same spin reorientation is not observed for LaMnAsO. LnMnAsO (Ln = La, Nd) are semiconducting [5, 6] and a sizeable negative magnetoresistance is observed below $T_{Mn}$. Magnetoresistance (MR) is defined as the change of electrical resistivity ρ in an applied magnetic field H, so that MR= (ρ(H)-ρ(0))/ρ(0). MR up to -24 % is observed at 200 K for LaMnAsO which is unprecedented for divalent $Mn^{2+}$. It has been postulated that the MR is a result of a reduction in quantum destructive interference upon application of a field [5].

Upon a small amount of substitution of $O^{2-}$ by $F^-$ in $NdMnAsO_{1-x}F_x$ (x = 0 – 0.08) the electronic properties change dramatically and colossal magnetoresistance (CMR) is observed at low temperature [8]. Colossal magnetoresistance is a rare phenomenon in which the electronic resistivity of a material can be decreased by orders of magnitude upon application of a magnetic field. CMR is well known in manganese oxides such as the perovskite $La_{1-x}Sr_xMnO_3$ [9, 10]. The CMR is observed below the spin reorientation transition in the antiferromagnetic oxypnictide $NdMnAsO_{1-x}F_x$ and is a result of competition between an antiferromagnetic insulating phase and a paramagnetic semiconductor upon application of a magnetic field [8]. In contrast there is little change to the magnetic properties upon substitution of $O^{2-}$ by $F^-$. Powder neutron diffraction data evidence antiferromagnetic order of the $Mn^{2+}$ moments below 356 (2) K [8], with the same magnetic structure as previously reported for NdMnAsO [5,6] so there is no appreciable change in $T_{Mn}$. The effects of H⁻ substitution have recently been explored in LaMnAsO$_{1-x}$H$_x$ ($x$= 0 – 0.73) [11]. At low doping levels ($x$ = 0.08) the materials remain semiconducting and very large magnetoresistances are observed at low temperature (MR$_{5T}$ (8 K) = -63 %). Upon increasing $x$ to 0.14, the LaMnAsO$_{1-x}$H$_x$ materials become metallic and ferromagnetic and the low temperature MR diminishes.

In this paper we report the electronic and magnetic properties of the series Nd$_{1-x}$Sr$_x$MnAsO ($x$ = 0, 0.05, 0.10). Hole doping of the [MnAs]⁻ layers results in metallic behaviour for $x \geq 0.05$ and positive magnetoresistance is observed at low temperature.

## II. EXPERIMENTAL

Nd$_{1-x}$Sr$_x$MnAsO ($x$ = 0, 0.05, 0.10) samples were prepared by solid state reaction of stoichiometric amounts of MnO$_2$, Mn and SrO powders (Aldrich 99.99 %) with pre-synthesised NdAs and MnAs precursors. All powders were ground under an inert atmosphere and pressed into pellets of 10 mm diameter. The pellets were heated at 1150°C for 48 hours in an evacuated quartz tube.

Powder X-ray diffraction patterns of Nd$_{1-x}$Sr$_x$MnAsO were collected using a Bruker D8 Advance diffractometer with twin Gobel mirrors and Cu Kα radiation. Data were collected over the range 10° < 2θ <100° with a step size of 0.02° and could be indexed on a tetragonal unit cell of space group *P4/nmm*, characteristic of the LnMAsO family (M=Fe, Ni, Co) [12].

Powder synchrotron X-ray diffraction patterns of Nd$_{1-x}$Sr$_x$MnAsO ($x$ = 0, 0.05 and 0.10) were recorded on the ESRF beamline ID31 at 290 K using a wavelength of 0.39996 Å. The sample was contained in a 0.5 mm diameter borosilicate glass capillary mounted on the axis of the diffractometer about which it was spinned at ~1 Hz to improve the powder averaging of the crystallites. Diffraction patterns were collected over the angular range 2 – 50° 2θ and rebinned to a constant step size of 0.002º for each scan. The high-angle parts of the pattern were scanned several times to improve the statistical quality of the data in these regions.

Powder neutron diffraction patterns were recorded on the high intensity diffractometer D20 at the ILL, Grenoble with a wavelength of 2.4188 Å. 1 g samples of $Nd_{1-x}Sr_xMnAsO$ ($x$ = 0.05, 0.10) were inserted into an 8 mm vanadium can and data were recorded at selected temperatures between 8 K and 360 K for $x$ = 0.05 and between 306 K and 360 K for $x$ = 0.10 with a collection time of 10 minutes per temperature.

### III. RESULTS AND DISCUSSION

#### A. Structural Trends

The powder X-ray diffraction patterns of $Nd_{1-x}Sr_xMnAsO$ show that samples with $x$ up to 0.10 are good quality. Minor impurity phases MnAs (~ 3 %), $Nd_2O_3$ (~2 %) and $MnO_2$ (~1 %) are observed for all samples. Any further increase in $x$ results in large quantities of secondary phases so that the solubility limit is 0.1. The synchrotron X-ray powder diffraction patterns of $Nd_{1-x}Sr_xMnAsO$ were fitted by the Rietveld method [13] using the GSAS program [14]. The backgrounds were fitted using linear interpolation and the peak shapes were modelled using a pseudo–Voigt function. The MnAs, $Nd_2O_3$ and $MnO_2$ impurities were also modelled giving the volume fractions displayed in Table 1. Our data confirm that $Nd_{1-x}Sr_xMnAsO$ crystallizes with the ZrCuSiAs type-structure (Fig. 1) and there is no change in symmetry from the *P4/nmm* space group upon doping with Sr (Table I). An excellent fit to this crystallographic model is obtained for all x (Figure 1). Rietveld refinement of the high resolution diffraction data confirms the successful substitution of $Sr^{2+}$ at the $Nd^{3+}$ site with the unconstrained occupancies refining to 0.050(2) and 0.099(2) (Table 1). The results also show an increase in both *a* and *c* cell parameters with increasing Sr doping (Figure 2) as expected considering the larger ionic radius of $Sr^{2+}$ (1.26 Å) compared to $Nd^{3+}$ (1.109 Å).The refinement results also show that both compounds are cation and anion stoichiometric and there is no evidence of anion disorder or cation disorder between Mn and Nd/Sr sites.

The variation of selected bond lengths and angles with *x* are displayed in Table 1; upon increasing *x* from 0 – 0.1 there are no clear trends in either the Nd/Sr-O or Mn-As bond lengths. The O-Nd-O bond angles and the Mn-As-Mn bond angles increase with x, which results in a reduction in the thickness of both the [MnAs]$^-$ (d$_{MnAs}$) and [NdO]$^+$ (d$_{NdO}$) layers. The overall consequence is an increase in the interlayer distance between Nd-O-Nd and As-Mn-As blocks with x, demonstrated by the increase in the Nd-As bond length (Table I). The interlayer distance also increases upon Sr doping in superconducting Pr$_{1-x}$Sr$_x$FeAsO [15].

B. Magnetic Structure

The magnetic susceptibilities of Nd$_{1-x}$Sr$_x$MnAsO samples (x = 0.00, 0.05 and 0.10) were measured between 5 K and 400 K on a Quantum Design SQUID magnetometer in an applied field of 100 Oe after zero-field cooling (ZFC). The susceptibility for Nd$_{1-x}$Sr$_x$MnAsO (x = 0.05, 0.10) is dominated by the ferromagnetic impurity MnAs with a broad ferromagnetic transition at 320 K. A sample of MnAs was prepared and its molar susceptibility measured and subtracted from the data as previously reported for Sr$_2$M$_3$As$_2$O$_2$ [16]. The low temperature susceptibility is shown in Figure 2. The Nd antiferromagnetic transition at ~ 20 K is apparent for all x. It was not possible to fully subtract the susceptibility of the MnAs impurity from the susceptibility data as the ferromagnetic transition of the MnAs impurity is much broader than the "synthesised" MnAs. It was therefore not possible to determine $T_{Mn}$ from the magnetic susceptibility data so that variable temperature neutron diffraction data were recorded in order to establish any change in $T_{Mn}$ upon increasing x in the Nd$_{1-x}$Sr$_x$MnAsO solid solutions.

NdMnAsO is a local moment antiferromagnet below $T_{Mn}$ = 359 K [7]. The Mn$^{2+}$ moments are aligned antiferromagnetically in the ab plane, but ferromagnetically along c. The moment is ordered parallel to the c axis and at 290 K the Mn$^{2+}$ moment refines to 2.35(2) μ$_B$ [5-7]. Neutron powder diffraction patterns of Nd$_{1-x}$Sr$_x$MnAsO evidence the (101) and (100) magnetic reflections below 342 K and 325 K for x = 0.05 and 0.10 respectively. These magnetic reflections can be indexed with a propagation vector k = (0, 0, 0), leading to identical magnetic and nuclear cells. Rietveld refinement of the magnetic structure reveals the same magnetic structure as NdMnAsO [5, 6]. The results demonstrate that substitution of Nd with Sr results in a reduction in $T_{Mn}$ from 359 K for $x$ = 0 to 325 K for $x$ = 0.1. There is no evidence of magnetic diffraction from the minor MnAs phase upon cooling ($T_c$ = 313 K [17]). The variation of the refined Mn moment with temperature is displayed in Figure 3 for both samples. A fit to $m=m_0(1-T/T_N)^\beta$ gives $m_0$= 3.6(1) $\mu_B$ and 3.4(1) $\mu_B$; $T_N$ = 342(2) K and 325 K; $\beta$ = 0.30(2) and 0.28(3). For $x$ = 0.05 an unusual variation of the Mn moment with temperature is observed. The variation of the Mn moment with temperature is very different from that reported for NdMnAsO [7] and NdMnAsO$_{1-x}$F$_x$ [8] where a smooth increase in the moment is observed upon cooling to 4 K.

Upon cooling below $T_{Mn}$ the Mn moment rapidly rises to 2.25(7) $\mu_B$ at 210 K, followed by a further gradual increase in moment to 2.83 $\mu_B$ at 24 K. A similar variation of the Mn moment has been reported for NdMnAsO [6, 7] and seems to be a general feature of these materials. However at 24 K the Mn moment of 2.83 $\mu_B$ is much smaller than that reported for NdMnAsO and NdMnAsO$_{1-x}$F$_x$ (3.60(1) $\mu_B$ and 3.75(4) $\mu_B$ respectively [6-8]). The moment is also much smaller than would be predicted from the fit to $m=m_0(1-T/T_N)^\beta$ (Fig. 3). This indicates that a significant component of the Mn moments remain disordered or paramagnetic down to 24 K. Below 24 K ($T_{Nd}$) the Nd spins order antiferromagnetically with moments aligned parallel to $a$. A change in the intensity of the magnetic Bragg peaks is observed below $T_{Nd}$ so that the (101) reflection is reduced compared to the (100) and an increase in intensity of the (102) peak is observed (Fig. 4). The variation of the Nd moment with temperature is shown in the inset to Figure 3. The antiferromagnetic order of the Nd spins induces a spin reorientation of the Mn moments into the basal plane as previously reported for NdMnAsO [6, 7] and NdMnAsO$_{1-x}$F$_x$ [8] (Fig. 4). However at $T_{Nd}$ the Mn moment abruptly increases from 2.83 $\mu_B$ to 3.64(6) $\mu_B$ and there is no further change upon cooling (Fig. 3). Recently it has been reported that there is competition between magnetic phases in electron doped NdMnAsO$_{1-}$ $_xF_x$ [8]. Upon applying a magnetic field, a second ordered phase transition from the antiferromagnetic phase to the paramagnetic phase is observed which suggests the presence of an antiferromagnetic instability. The presence of a significant component of disordered Mn moments in $Nd_{0.95}Sr_{0.05}MnAsO$ suggests that similar behaviour is observed for the hole doped materials so that some of the Mn spins remain paramagnetic, even in H = 0 T. Below $T_{Nd}$, the magnetic coupling between Mn and Nd is sufficient to induce antiferromagnetic order of the residual paramagnetic spins so that at 4 K the Mn moment is 3.62 (6) $\mu_B$ which is comparable to that reported for NdMnAsO and $NdMnAsO_{1-x}F_x$ (3.60(1) $\mu_B$ and 3.75(4) $\mu_B$ respectively).

## C. Electronic properties

The electronic transport properties of sintered ceramic bars of $Nd_{1-x}Sr_xMnAsO$ were investigated using a Quantum Design Physical Property Measurement System (PPMS). The resistivities of the bars were measured, using the 4-point probe technique, as a function of temperature in the range 5 K < $T$ < 300 K in both zero applied field and with a 7 T applied field. The variation of resistivity with temperature for $x$ = 0.05 and $x$ = 0.10 are displayed in Figure 4. Both materials appear metallic with 300 K resistivites of 0.002 $\Omega$.Cm and 0.0012 $\Omega$.Cm and residual resistance ratios ($\rho_{300K}/\rho_{5K}$) of 6.3 and 5.8 for $x$ = 0.05 and $x$ = 0.10 respectively. From 5 K to 60 K, Fermi liquid $T^2$ behaviour is observed for both materials which suggests that hole-hole scattering is the dominant contribution to the resistivity at low temperature. A crossover to linear behaviour is observed at higher temperature. NdMnAsO is semiconducting (Fig. 5) [5, 6] and the resistivity at room temperature is ~ 4300 $\Omega$.Cm. Upon substitution of $Nd^{3+}$ with $Sr^{2+}$ the electronic properties change abruptly so that metallic behaviour is observed for $x \geq 0.05$ as a result of hole doping of the $[MnAs]^-$ layer. A similar result is observed for $La_{1-x}Sr_xMnAsO$ [18]. The metallisation of $Ln$MnAsO oxypnictides appears to be unsymmetrical to hole and electron doping [18], for example the transition from the Mott insulator to the metal state occurs between $x$ = 0.17 and 0.2 in electron doped $SmMnAsO_{1-x}$ [19]. The results demonstrate that $Nd_{1-x}Sr_xMnAsO$ is driven metallic for $x \geq 0.05$ whilst retaining the same antiferromagnetic structure of the Mn spins. $T_{Mn}$ reduces slightly from 359 K for $x = 0$ to 325 K for $x = 0.1$ which suggests that $Nd_{1-x}Sr_xMnAsO$ remains an antiferromagnetic local moment metal for $x \geq 0.05$. Similar results have been reported for hole doped $Ba_{1-x}K_xMn_2As_2$ [20] where it is suggested that the electronic conduction is mainly due to itinerant As 4p holes.

Surprisingly the low temperature antiferromagnetic ordering transition of the Nd spins and subsequent spin reorientation of the Mn moments is not evident in the resistivity data. The neutron diffraction measurements evidence an unusual variation of the Mn moment with temperature so that below $T_{Mn}$ there is an initial rapid increase in the Mn moment to 2.25(7) $\mu_B$ at 210 K consistent with a second order phase transition. This is followed by a further gradual increase in the moment, reaching 2.83 $\mu_B$ at 24 K, which clearly manifests in the electronic resistivity of both materials below ~ 230 K where a change in slope is detected (as indicated in Figure 4) that most likely results from a change in magnetic scattering of the charge carriers.

Figure 6 shows the variation of the magnetoresistance ($MR_{7T}$) with temperature for $Nd_{1-x}Sr_xMnAsO$. There is no evidence of the large negative magnetoresistance observed for NdMnAsO between 150 – 390 K, but at low temperature a positive MR is observed for both Sr doped samples. This is in stark contrast to the change in magnetotransport upon electron doping where a colossal magnetoresistance of up to -95% is observed at 3 K for $NdMnAsO_{1-x}F_x$ [8]. The positive MR reaches a maximum at ~ 15 K which is not associated with any of the magnetic transitions observed for $Nd_{1-x}Sr_xMnAsO$. The positive MR could arise due to a change in mobility associated with a change in band structure upon cooling as reported for La(Fe, Ru)AsO [21]. Electronic structure calculations of hole doped LnMnAsO oxypnictides are clearly warranted to explore this further.

Finally we note that MnAs is observed as a minor impurity phase for all $Nd_{1-x}Sr_xMnAsO$ ($x$= 0.00, 0.05, 0.10) phases. MnAs also exhibits metallic behaviour but does not exhibit any anomalies in the resistivity between 10 –300 K and also does not exhibit low temperature positive MR. Hence the electronic results reported above arise from hole doping of the [MnAs]$^-$ slab.

In summary we have shown that it is possible to substitute up to 10 % Sr for Nd in NdMnAsO. There is a remarkable change in the electronic properties so that metallic behaviour is observed for $x \geq 0.05$. In contrast there is little change to the magnetic transition temperature, $T_{Mn}$ is reduced from 359 K for $x = 0$ to 325 K for $x = 0.1$ so that the Sr doped materials are local moment metals. A positive magnetoresistance is observed at low temperature in Nd$_{1-x}$Sr$_x$MnAsO ($x = 0.05$ and 0.1) which illustrates that multiple MR mechanisms are possible in *Ln*MnAsO oxypnictides.

## IV    ACKNOWLEDGEMENTS

We acknowledge the UK EPSRC for financial support and STFC-GB for provision of beamtime at ILL and ESRF.

Table I: Refreshed cell parameters, agreement factors, atomic parameters and selected bond lengths and angles for $Nd_{1-x}Sr_xMnAsO$ from Rietveld fits against ID31 synchrotron X-ray powder diffraction data at 290 K. Nd/Sr and As are at 2c (¼, ¼, z), Mn at 2b (¾, ¼, ½) and O at 2a (¼, ¾, 0).

| Atom | | x 0.00 | 0.05 | 0.10 |
|---|---|---|---|---|
| Nd/Sr | Nd occupancy | 0.998(1) | 0.951(1) | 0.90(1) |
| | Sr occupancy | 0.000 | 0.050(2) | 0.099(2) |
| | z | 0.12972(2) | 0.12907(4) | 0.12869(4) |
| | $U_{iso}$ (Å$^2$) | 0.00400(4) | 0.00367(7) | 0.00112(7) |
| Mn | Occupancy | 0.993(2) | 0.995(2) | 0.995(2) |
| | $U_{iso}$ (Å$^2$) | 0.0066(1) | 0.0070(2) | 0.0025(2) |
| As | Occupancy | 1.005(2) | 1.003(2) | 1.003(2) |
| | z | 0.673492(5) | 0.67244(7) | 0.67209(7) |
| | $U_{iso}$ (Å$^2$) | 0.00573(8) | 0.0056(1) | 0.0021(1) |
| O | Occupancy | 0.995(7) | 0.997(8) | 0.997(8) |
| | $U_{iso}$ (Å$^2$) | 0.0066(5) | 0.0071(9) | 0.0040(9) |
| | $a$ (Å) | 4.051664(3) | 4.053803(9) | 4.056151(8) |
| | $c$ (Å) | 8.90676(1) | 8.91573(4) | 8.92874(3) |
| | Cell volume (Å)$^3$ | 146.213(1) | 146.515(1) | 146.899(1) |
| | $\chi^2$ (%) | 3.83 | 5.45 | 4.45 |
| | $R_{WP}$ (%) | 10.80 | 12.55 | 12.57 |
| | $R_P$ (%) | 7.91 | 8.72 | 10.57 |
| | Nd/Sr-O (Å) x 4 | 2.3321(1) | 2.3307(2) | 2.3309(2) |
| | Mn-As (Å) x 4 | 2.5475(2) | 2.5440(4) | 2.5444(4) |
| | Mn-Mn (Å) x 4 | 2.86649(1) | 2.86647(1) | 2.86813(1) |
| | Nd-As (Å) x 4 | 3.3589(3) | 3.3688(4) | 3.3749(4) |
| | O-Nd-O (1) (°) x 2 | 120.60(1) | 120.82(2) | 120.92(2) |

| | | | |
|---|---|---|---|
| O-Nd-O (2) (°) x 4 | 75.792(5) | 75.892(7) | 75.936(6) |
| Mn-As-Mn (1) (°) x 2 | 105.35(2) | 105.63(2) | 105.70(2) |
| Mn-As-Mn (2) (°) x 4 | 68.43(1) | 68.58(1) | 68.61(11) |
| MnAs layer (Å) | 3.0905(1) | 3.0749(1) | 3.0731(1) |
| NdO layer (Å) | 2.3108(1) | 2.3015(1) | 2.2981(1) |
| MnAs (Vol. %) | 2.0(1) | 2.9(1) | 2.8(1) |
| Nd$_2$O$_3$ (Vol. %) | 1.1(1) | 1.5(1) | 1.4(1) |
| MnO (Vol. %) | 0.5(1) | 0.9(1) | 0.8(1) |

Figure Captions

Fig. 1 (color online) a) Crystal structure of $Nd_{1-x}Sr_xMnAsO$; the O-Nd-O and Fe-As-Fe bond angles are displayed below b) Rietveld refinement fit to the 290 K ID31 synchrotron X-ray powder diffraction pattern of $Nd_{0.9}Sr_{0.1}MnAsO$. Tick marks represent reflection positions for $Nd_{0.95}Sr_{0.05}MnAsO$ and MnAs, $MnO_2$ and $Nd_2O_3$ minor impurity phases from bottom to top respectively.

Fig. 2 (color online) Zero field cooled magnetic susceptibility of $Nd_{1-x}Sr_xMnAsO$ versus temperature, recorded in a 100 Oe magnetic field.

Fig. 3 (color online) The temperature variation of the Mn moment for $Nd_{0.95}Sr_{0.05}MnAsO$ for moments aligned parallel to *c* (filled circles) and aligned parallel to the basal plane (open circles) and for $Nd_{0.9}Sr_{0.1}MnAsO$ (crosses). An unusual variation of the Mn moment with temperature is evidenced. Below $T_{Mn}$ the Mn moment rapidly rises to 2.25(7) $\mu_B$ at 210 K, followed by a further gradual increase in moment to 2.83 $\mu_B$ at 24 K as indicated by the arrow. The solid lines shows the fit to $m=m_0(1-T/T_N)^\beta$. Antiferromagnetic order of the Nd spins below 20 K appears to induce magnetic order of the residual paramagnetic Mn spins. The temperature variation of the Nd and Mn moments is shown in the inset for *x* = 0.05 between 8 and 70 K.

Fig. 4 A portion of the 8 K D20 neutron diffraction pattern for $Nd_{0.95}Sr_{0.05}MnAsO$ showing the magnetic diffraction peaks observed below $T_{Nd}$. The difference curve to the Rietveld refinement to the neutron data is shown below. The inset shows the magnetic structure above $T_{Nd}$ (left) and below $T_{Nd}$ (right).

Fig. 5 (color online) The resistivity of polycrystalline $Nd_{1-x}Sr_xMnAsO$ (*x* = 0.05 and 0.10) measured in zero field. Fermi liquid $T^2$ behaviour is observed at low temperature and two linear regions are observed upon warming as indicated by the arrows. The inset shows the variation of resistivity with temperature for NdMnAsO.

Fig. 6 (color online) Variation of 7 T MR with temperature for $Nd_{1-x}Sr_xMnAsO$.

Figure 1

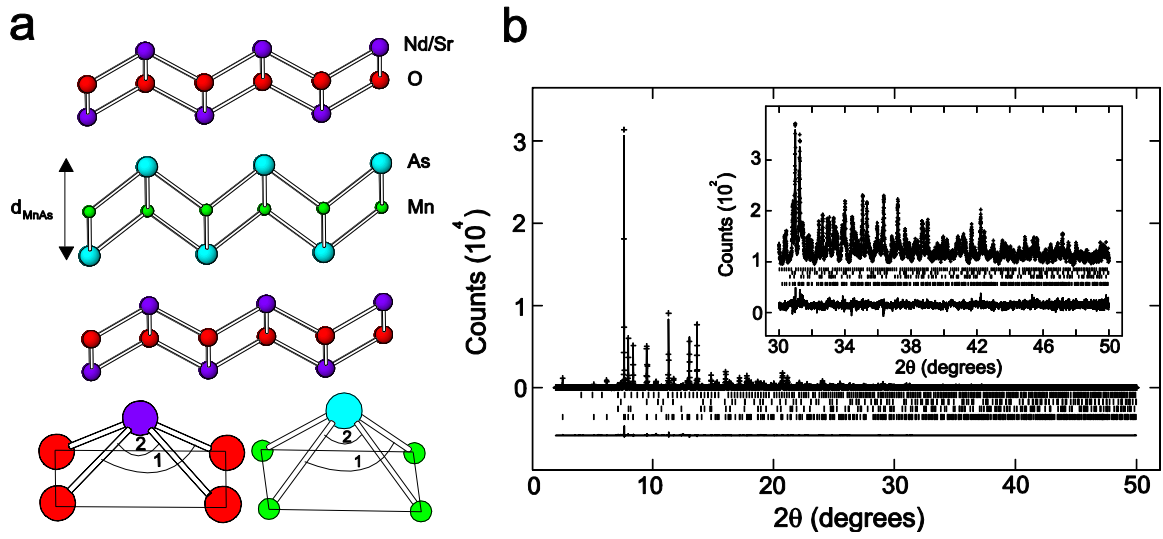

Figure 2

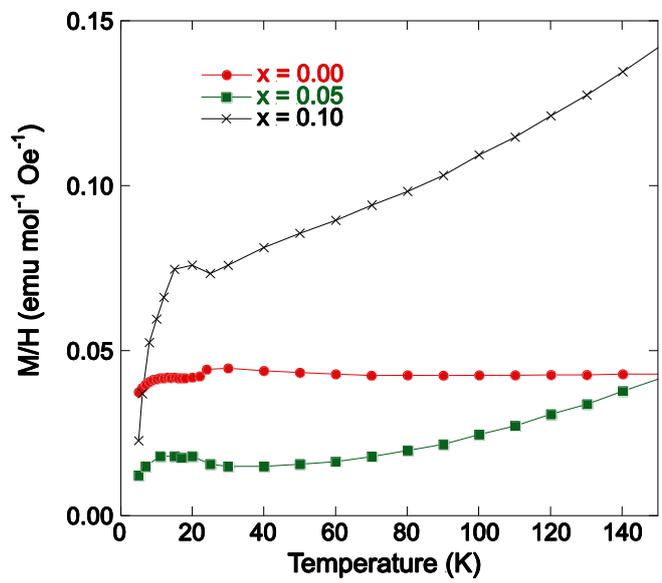

Figure 3

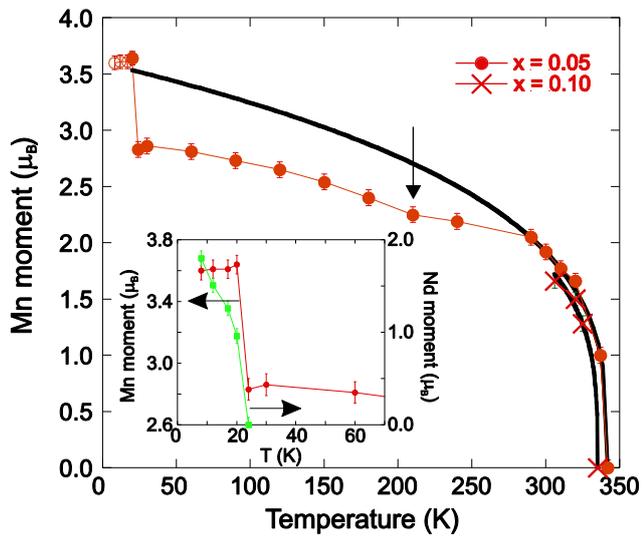

Figure 4

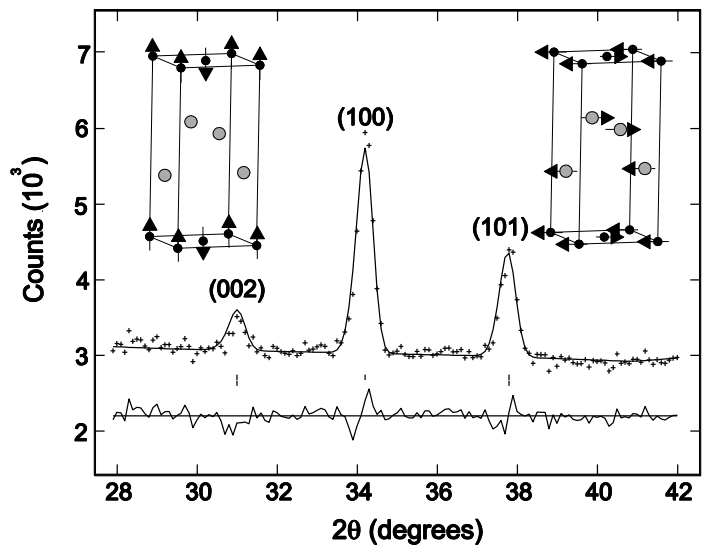

Figure 5

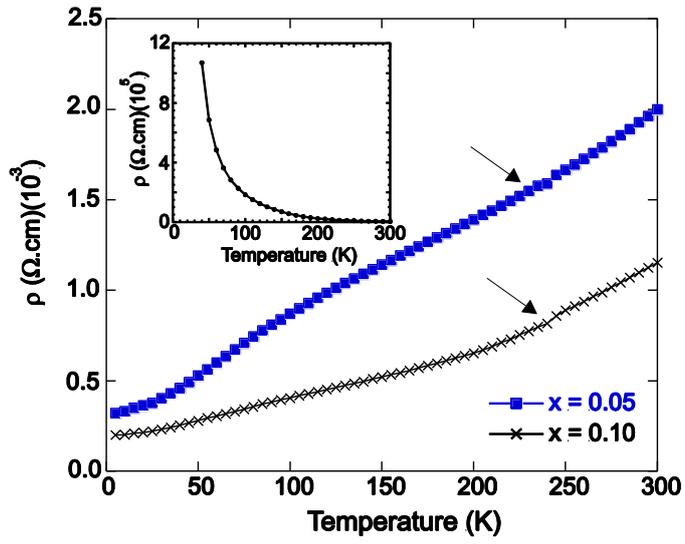

Figure 6

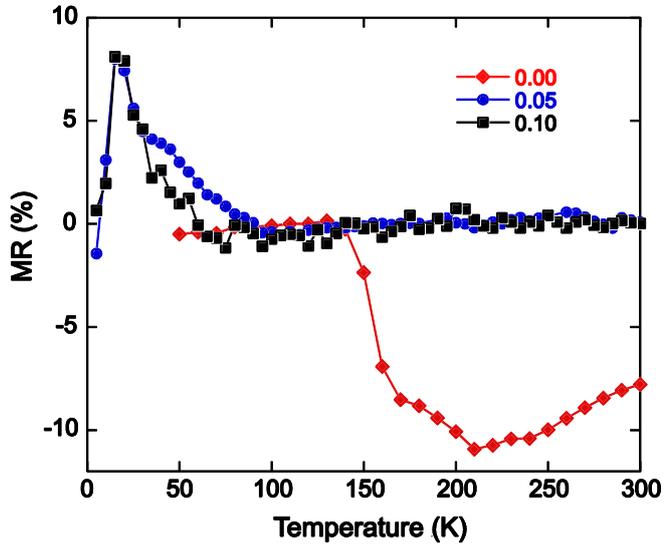